\documentclass[pra,aps,twocolumn,showpacs]{revtex4}
\usepackage{epsfig}
\usepackage{amsmath}
\begin{document}

\title{Josephson Oscillation and Transition to Self-Trapping  for Bose-Einstein-Condensates in a Triple-Well Trap}
\author{Bin Liu$^{1,2}$, Li-Bin Fu$^{2}$, Shi-Ping Yang$^{1}$, and Jie Liu$^{2*}$}
 \affiliation{1. College of Physics and Information
 Engineering, Hebei Normal University, 050016 Shijiazhuang, China  \\
2. Institute of Applied Physics and
Computational Mathematics, P.O. Box 8009
(28), 100088 Beijing, China
}
\begin{abstract}
We investigate the tunnelling dynamics of
Bose-Einstein-Condensates(BECs)  in a symmetric as well as in a
tilted triple-well trap within the framework of mean-field
treatment. The eigenenergies  as the functions of the zero-point
energy difference between the  tilted wells show a striking
entangled star structure when the atomic interaction is large. We
then achieve insight into the oscillation solutions around the
corresponding eigenstates and observe several new types of Josephson
oscillations. With increasing the atomic interaction, the
Josephson-type oscillation is blocked and the self-trapping solution
emerges. The condensates are self-trapped either in one well or in
two wells but no scaling-law is observed near transition points. In
particular, we find that the transition from the Josephson-type
oscillation to the self-trapping is accompanied with some irregular
regime where tunnelling dynamics is dominated by chaos. The above
analysis is facilitated with the help of the Poicar\'{e} section
method that visualizes the motions of BECs in a reduced phase plane.
\end{abstract}
 \maketitle

\section{Introduction}

Since the first realization of dilute degenerate atomic gases in
1995, a new epoch for studying the dynamical property of
Bose-Einstein Condensates (BECs) comes\cite{first}. For the dilute
degenerate gases, essential dynamical property  is included in the
Gross-Pitaevskii equation (GPE)\cite{gpe}. The nonlinearity,
originated from the interatomic interaction, is included in the
equation through a mean field term proportional to condensate
density. Previously, several authors investigated the dynamics of
GPE for a double-well potential in a two-mode
approximation\cite{twomode,niu,stat,jose,lzt,leggett,add}. Novel
features were found, e.g., the emergence of new nonlinear
stationary states\cite{stat} and a variety of new crossing
scenarios\cite{add}, nonzero adiabatic tunnelling
probability\cite{niu,lzt}, etc, to name only a few. Among these
findings, nonlinear Josephson oscillation and self-trapping
phenomenon are of most interest. As well be known, for single
particle in a symmetric double-well, the tunnelling dynamics is
determined   by  the tunnelling splitting of two nearly degenerate
eigen-states and tunnelling time or quantum oscillation period is
inversely proportional to the energy splitting\cite{qm}. When
atomic interaction emerges, the tunnelling between two-well is
also observed, termed as nonlinear Josephson
oscillation\cite{twomode,jose,leggett}. However, in this case, the
oscillation period  sensitively depends on the initial state but
has little relation to the difference between the eigenenergies.
More interestingly, with increasing the atomic interaction
further(even it is repulsive), the Josephson oscillation between
two wells is completely blocked, the BECs atoms in a symmetric
double-well potential shows a highly asymmetric distribution as if
most atoms are trapped in one well\cite{twomode}. This somehow
counterintuitive phenomenon is termed as self-trapping and has
been observed in lab recently\cite{exper}.
\begin{figure}[b]
\centering \rotatebox{0}{\resizebox *{8.0cm}{5.0cm}
{\includegraphics {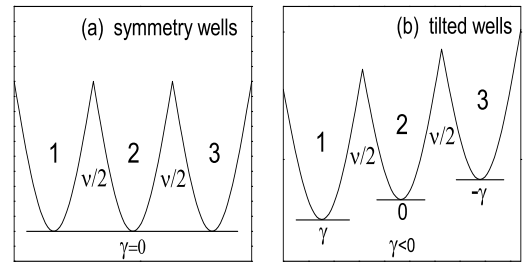}}}
\caption{The schematic sketch of
our model. (a) The symmetric case($\gamma=0$); (b) The asymmetric
case, ($\gamma,0,-\gamma$) is the zero-point energy in each well
respectively.} \label{potential}
\end{figure}

In the present paper, we  extend to investigate tunnelling
dynamics for BECs in a triple-well system\cite{tri,ple}
(schematically sketched as Fig.\ref{potential}), want to know how
the nonlinear Josephson oscillation and self-trapping behave in
this simplest multi-well system. This extension is not trivial
because quantum tunnelling may happen between several wells
simultaneously so that we expect that the tunnelling dynamics in
the triple-well will show more interesting behavior. Moreover, the
study of triple-well system will provide a bridge between the
simple double-well and the multi-well,  helping us understand  the
'self-localized' phenomenon of BECs in the optical
lattice\cite{bbw}.

Technically, to investigate the dynamics of triple-well systems we
resort to the Poicar\'{e} section method\cite{ps} that visualizes
the motion of BECs in a reduced two dimensional phase plane. For
our triple-well system, ignoring a total phase the dynamics is
governed by a Hamiltonian with two freedoms. Its phase space is
four-dimensional. However,  the  motions in a high-dimensional
phase space ( in our case, it is 4D) are difficult to trace. With
using the Poicar\'{e} section, we can investigate the motions of
BEC in a reduced 2D phase plane.

Our paper is organized as follows. In Sec.II we introduce our model
and show the unusual structure of the eigen-energies. In Sec.III we
investigate nonlinear Josephson oscillations  with using
Poincar\'{e} section method and demonstrate diverse types of the
oscillations for BECs. In Sec.IV we investigate the transition from
the Josephson oscillation to self-trapping of BECs in one-well as
well as in two-well and show an irregular regime characterized by
chaos. Our discussions are extended to tilted triple-well system in
Sec.V. Final section is our discussions and conclusions.

\section{Model}
For the triple-well system and under mean field approximation,  the
wave function $\Psi (r,t)$ of GPE is the superposition of three wave
functions describing the condensate in each trap, i.e.,
\begin{equation}
\Psi (r,t)=a _{1}(t)\phi _{1}(r)+a _{2}(t)\phi _{2}(r)+a _{3}(t)\phi
_{3}(r).
\end{equation}

Then the  triple-well system is described by a dimensionless
Schr\"{o}dinger equation,

\begin{eqnarray}
i \frac{d}{dt}
\left( \begin{array}{c}
 a_1 \\
 a_2\\
 a_3
\end{array} \right)
=\hat{H}
\left( \begin{array}{c}
a_1 \\
 a_2\\
 a_3
\end{array} \right),
\label{triple}
\end{eqnarray}
with the Hamiltonian
\begin{eqnarray}
\hat{H}=\left(
\begin{array}{ccc}
   \gamma+c|a_1|^2 & -\frac{v}{2} & 0 \\
 -\frac{v}{2} & c|a_2|^2 & -\frac{v}{2} \\
 0 & -\frac{v}{2} & -\gamma+c|a_3|^2
\end{array}
\right).
\label{hamil2}
\end{eqnarray}
The total probability $\left| a_{1}\right| ^{2}+\left|
a_{2}\right|^{2}+\left| a_{3}\right| ^{2}$ is conserved and is set
to be unit. $c$ is the mean field parameter denoting the atomic
interaction and $v$ is the coupling parameter, $\gamma$ is the
zero-point energy of the wells. The schematic sketch of the model is
shown in Fig.\ref{potential}. In our following discussions, we focus
on the case of repulsive interaction between atoms, i.e., $c>0$.

 With ignoring a total phase, the dynamics of  the above
three-level quantum system can be depicted by  a classical
Hamiltonian  of two-degree freedom\cite{tri}. Let us set that,
$n_1=|a_1|^2,n_2=|a_2|^2,n_3=|a_3|^2,\theta_1=\arg a_1-\arg
a_2,\theta_3=\arg a_3-\arg a_2$, use the constraint $n_1+n_2+n_3=1$,
we can get the classical Josephson Hamiltonian,
\begin{eqnarray}
 \mathcal{H}&=&\gamma(n_1-n_3)+\frac{1}{2}c\left(n_1^2+n_3^2+\left(1-n_1-n_3\right)^2\right) \nonumber \\
&&-v\sqrt{1-n_1-n_3}\left(\sqrt{n_1}
\cos\theta_1+\sqrt{n_3}\cos\theta _3\right), \label{joseham}
\end{eqnarray}
and the corresponding canonical equations

\begin{subequations}
\begin{align}
\frac{dn_1}{dt} &= -v \sin \left(\theta _1\right) \sqrt{n_1} \sqrt{1-n_1-n_3}  ,\\
\frac{d \theta_1}{dt} &= \gamma +\frac{1}{2} c \left(2 n_1-2
   \left(1-n_1-n_3\right)\right)                                            \nonumber\\
 &  -\frac{v \cos \left(\theta
   _1\right) \sqrt{1-n_1-n_3}}{2 \sqrt{n_1}}                                \nonumber\\
 &  +\frac{v
   \left(\sqrt{n_1} \cos \left(\theta _1\right)+\cos \left(\theta
   _3\right) \sqrt{n_3}\right)}{2 \sqrt{1-n_1-n_3}}                      ,       \\
 \frac{dn_3}{dt} &= -v \sin \left(\theta _3\right) \sqrt{1-n_1-n_3} \sqrt{n_3} , \\
\frac{d \theta_3}{dt} &=-\gamma +\frac{1}{2} c \left(2 n_3-2
   \left(1-n_1-n_3\right)\right)                                            \nonumber\\
 &  +\frac{v \left(\sqrt{n_1} \cos
   \left(\theta _1\right)+\cos \left(\theta _3\right)
   \sqrt{n_3}\right)}{2 \sqrt{1-n_1-n_3}}                                   \nonumber\\
 &  -\frac{v \cos
   \left(\theta _3\right) \sqrt{1-n_1-n_3}}{2 \sqrt{n_3}}.
\label{cano}
\end{align}
\end{subequations}

\begin{figure}[t]
\centering
\includegraphics[width=0.5\textwidth]{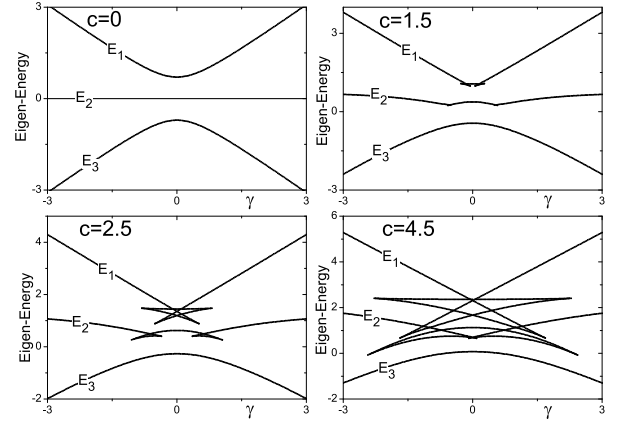}
\caption{The eigen-energy levels for different interaction strength.
We have set $v=1$.} \label{level}
\end{figure}

The fixed point or minimum energy point of the classical
Hamiltonian system(\ref{joseham}) corresponds to the eigen-state of quantum
system\cite{clas,lzt}. To derive the analytical expressions of
these fixed points is difficult, however, numerically, we can
readily obtain them with exploiting the Mathematica
Software\cite{wolf}. We plot the eigenenergies as the function of
the zero-point energy bias in Fig.2, they show unusual entangled
star structure for the strong nonlinearity.

For the weak interactions, the eigen-energy levels is very similar
to linear case($c=0$). With increasing the nonlinearity(i.e.,
c=1.5), topological structure of the upper level changes: two small
loops emerge.
 When the interaction is stronger(i.e., c=2.5), the two
loops will collide and form a star structure, and then two more
loops will emerge at the middle level $E_2$. For still stronger
interaction (i.e., c=4.5), the star structure of the upper level
entangles with the star structure in the  lower level. However, we
can still distinguish these levels because they have different
relative phases. In fact, levels labeled by $E_1$ have relative
phases $(\theta_1=\pi, \theta_3=\pi)$, levels labeled by  $E_2$
have relative phases $(\theta_1=\pi, \theta_3=0)$ or $(\theta_1=0,
\theta_3=\pi)$, levels labeled by $E_3$ have relative phases
$(\theta_1=0, \theta_3=0)$.

The relation between the chemical potential and the above energy
\begin{eqnarray}
E=\mu-\frac{c}{2}(|a_1|^4+|a_2|^4+|a_3|^4), \label{mu}
\end{eqnarray}
where $\mu$ denotes the chemical potential defined as
$\langle \Psi|H|\Psi \rangle$.

In the above calculations and henceforth, for convenience we set
the coupling parameter as unit, i.e., $v=1$.

\begin{figure}[b]
\centering
\includegraphics[width=0.5\textwidth]{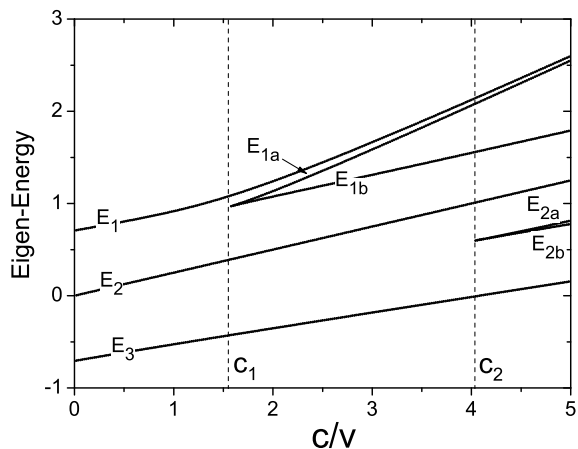}
\caption{When $\gamma=0$, the energy levels vary with the interaction strength $c/v$, where we
set $v=1$.}
\label{eigen}
\end{figure}

\section{The Symmetric Triple-Trap Case, $\gamma=0$}

Firstly we focus on the symmetric case, i.e.,
 $\gamma=0$.
The dependence of the energy levels on the interaction strength is
exposed by Fig.3. For the interaction strength is less than a
critical value $c_1=1.56$, the level structure is similar to its
linear counterpart except for some positive shifts on the energy
values. For $c>c_{1}$, two more levels labeled as $(E_{1a},
E_{1b})$ emerge. Actually, they correspond to  the star structure
of upper level $E_1$ in Fig.\ref{level} and have relative phases
of $(\theta_1=\pi, \theta_3=\pi)$. When the interaction strength
is still stronger and exceeds  the second critical value
$c_2=4.06$, the other two more energy levels labeled as $(E_{2a},
E_{2b})$ emerge. They correspond to  the star structure of
mid-level $E_2$ in Fig.\ref{level} and have the same relative
phases as that of  $E_2$.

The stability of the corresponding eigen-states can be evaluated
by the eigen-values  of the Jacobian of the classical Josephson
Hamiltonian (\ref{joseham}).
\begin{equation}
\mathcal{J}=\left(
\begin{array}{cccc}
-\frac{\partial ^{2}\mathcal{H}}{\partial n_1\partial \theta _{1}}   &
-\frac{\partial ^{2}\mathcal{H}}{\partial \theta _{1}^{2}}
& -\frac{\partial ^{2}\mathcal{H}}{\partial n_3\partial \theta _{1}} & -\frac{\partial ^{2}\mathcal{H}}{\partial \theta_3\partial \theta _{1}}\\
\frac{\partial ^{2}\mathcal{H}}{\partial n_1^{2}} & \frac{\partial
^{2}\mathcal{H}}{\partial \theta _{1}\partial n_1}
& \frac{\partial ^{2}\mathcal{H}}{\partial n_3\partial n_1}& \frac{\partial ^{2}\mathcal{H}}{\partial \theta_3\partial n_1}\\
 -\frac{\partial ^{2}\mathcal{H}}{\partial n_1\partial \theta_3} & -\frac{\partial ^{2}\mathcal{H}}{\partial \theta _{1}\partial \theta_3}
 & -\frac{\partial ^{2}\mathcal{H}}{\partial n_3\partial \theta_3}& -\frac{\partial ^{2}\mathcal{H}}{\partial \theta_3^{2}}\\
 \frac{\partial ^{2}\mathcal{H}}{\partial n_1\partial n_3}& \frac{\partial ^{2}\mathcal{H}}{\partial \theta _{1}\partial n_3}
 & \frac{\partial ^{2}\mathcal{H}}{\partial n_3^{2}} & \frac{\partial ^{2}\mathcal{H}}{\partial \theta_3\partial n_3}
\end{array}
\right).
\label{jac}
\end{equation}

The eigen-values  of the above Jacobian have their correspondence of
the Bogoliubov excitation spectrum of BECs . Pure imaginary values
indicates to  a stable  BECs state, whereas emergence of real values
implies the unstable for BECs and lead to a rapid production of the
Bogoliubov quasi-particles\cite{adda}. From calculating the above
Jacobian matrix and making diagonalization we know that, the states
corresponding to  level $E_{1b}$ and level $E_{2b}$ are unstable,
others are  stable.

\subsection{Linear Josephson Oscillation Solution}

\begin{figure}[b]
\centering
\includegraphics[width=0.5\textwidth]{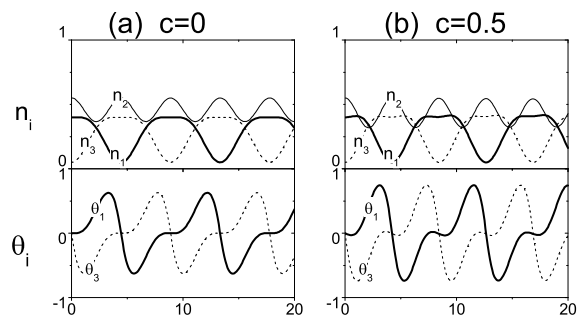}
\caption{The evolutions of $n_1$(heavy line), $n_2$(thin line),
$n_3$(dashed line) and $ \theta_1$(heavy line), $\theta_3$(dashed
line) in symmetric wells for linear case(a) and weak interactions
case(b), with the same initial values
($n_1=0.4,n_3=0.05,\theta_1=0,\theta_3=0$).} \label{alla}
\end{figure}

For the linear Josephson oscillation, i.e., $c=0$, the system is
analytically solvable. The solutions of $(a_1,a_2,a_3)$ are

\begin{subequations}
\begin{align}
a_1 &= C_2 \cos (\frac{v}{\sqrt{2}}t+C_3)+C_1                   ,       \\
a_2 &= C_4 \cos (\frac{v}{\sqrt{2}}t+C_5)                         ,     \\
a_3 &= -(C_2 \cos (\frac{v}{\sqrt{2}}(t+\frac{T}{2})+C_3)+C_1).
\end{align}
\label{solu}
\end{subequations}
Where $C_i$ are parameter determined by initial conditions

\begin{subequations}
\begin{align}
& C_2 \cos C_3 +C_1=a_1(0)                                       ,      \\
& C_4 \cos C_5 =a_2(0)                                            ,     \\
& C_2 \cos C_3 -C_1=a_3(0)                                         ,    \\
& -\frac{v}{\sqrt{2}} C_2 \sin C_3 =i \frac{v}{2} a_2(0)           ,    \\
& -\frac{v}{\sqrt{2}} C_4 \sin C_5 =i \frac{v}{2} (a_1(0)+a_3(0))  ,
\end{align}
\label{initial}
\end{subequations}

and the constraint
\begin{eqnarray}
|a_1(0)|^2+|a_2(0)|^2+|a_3(0)|^2=1.
\label{const}
\end{eqnarray}

From the above explicit expressions, we see that  $a_1,a_2$ and
$a_3$  vary with respect to time periodically. They share a common
period that is inversely proportional to the coupling parameter,
i.e., $T=2\sqrt{2}\pi/v$. Actually, the frequency is just the bias
between eigen-energy levels.  With initial conditions, the
coefficients $C_i$ will be fixed by using (\ref{initial}). In our case,
$C_1\ne 0$, so the period of population $n_1$ and $n_3$ is twice
the period of $n_2$, and compared with $n_1$,  variable $n_3$
has a phase delay of half-period. The above analysis is confirmed
by our numerical simulations as shown in Fig.\ref{alla}(a).

\begin{figure}[t]
\centering
\includegraphics[width=0.5\textwidth]{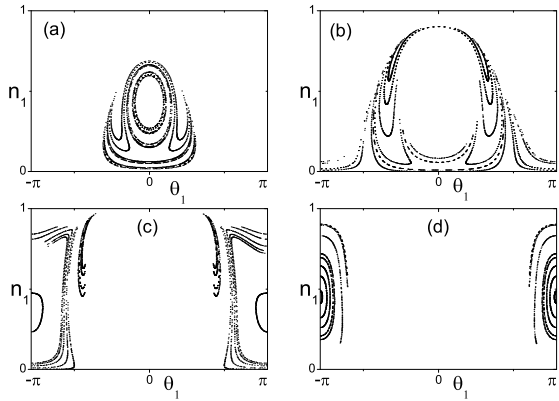}
\caption{The Poincar\'{e} section at $\theta_3=0$ for $c/v=0.5$ with different energy $E$. (a)$E=-0.4$,
(b)$E=-0.1$, (c)$E=0.2$, (d)$E=0.5$.}
\label{poinsm}
\end{figure}

\subsection{Weak Interaction cases, $c<c_1$}

The dynamics of this two-freedom system could be visualized from
Poincar\'{e} section\cite{ps}. We do this by solving the canonical
equations (\ref{cano}) numerically and then plotting $\theta_1$ and
$n_1$ at each time that $\theta_2=0$ and $\dot{\theta_2}<0$. Notice
the total energy is conserved,  therefore the Poincar\'{e} sections
consists of a picture panel where each picture corresponds a fixed
energy.

For the linear case, all of motions share a common period exactly,
and the Poincar\'{e} section is some isolated points. With weak
interactions, the periodicity will be destroyed, and the motions
become periodic or quasi-periodic, corresponding  Poincar\'{e}
section is plotted in Fig.\ref{poinsm}, where we see the section
plane is full of stable islands.
 However, in this case the motion is similar to the linear case if
they have the same initial values, as shown in Fig.\ref{alla}(a),\ref{alla}(b).

\subsection{Strong Interaction Cases, $c>c_2$}

\begin{figure}[t]
\centering
\includegraphics[width=0.5\textwidth]{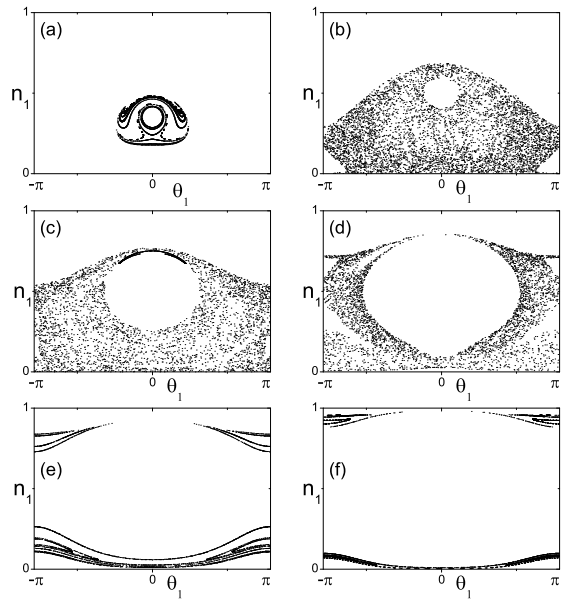}
\caption{The poincar\'{e} section at $\theta_3=0$ for $c/v=5$ with different energy $E$. (a)$E=0.3$,
(b)$E=0.8$, (c)$E=1.1$, (d)$E=1.5$, (e)$E=1.9$, (f)$E=2.3$.}
\label{poinbi}
\end{figure}

\begin{figure*}[t]
\centering
\includegraphics[width=0.95\textwidth]{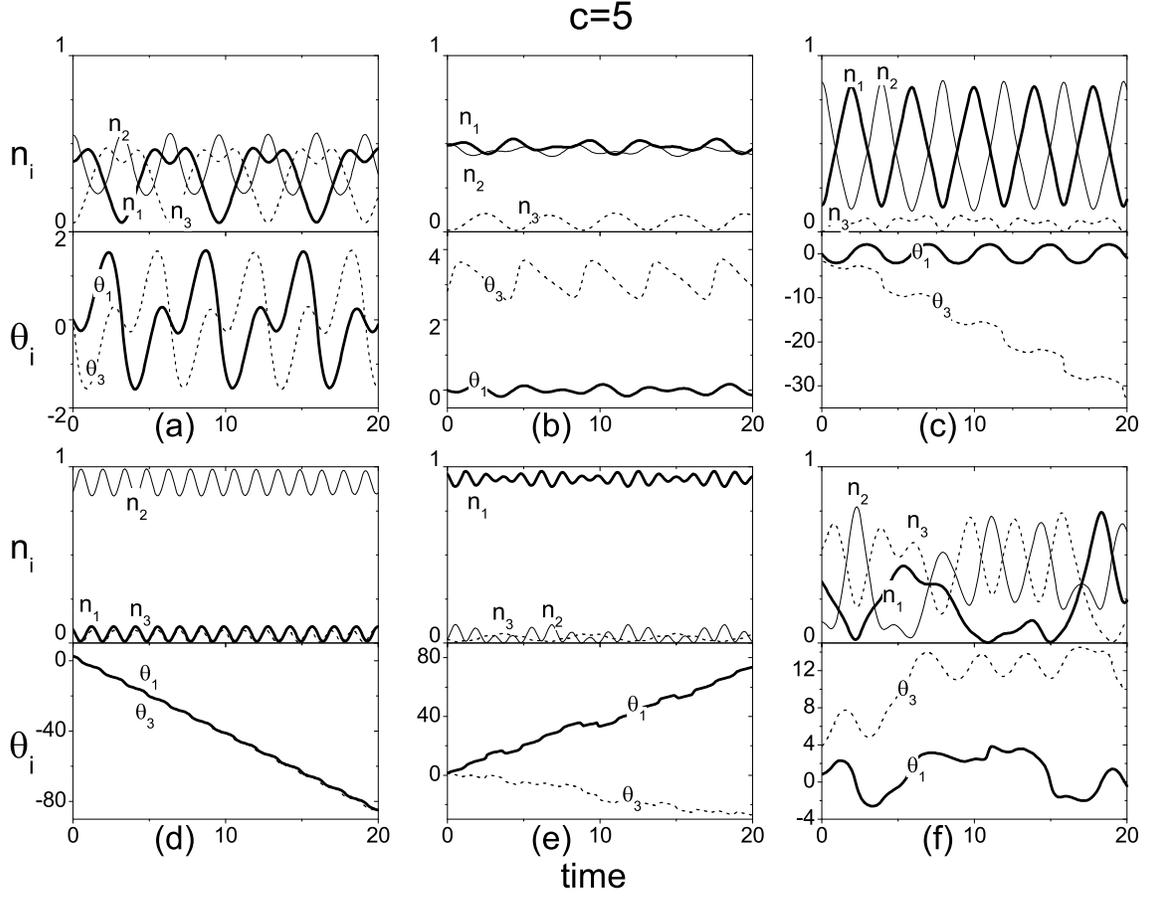}
\caption{The evolutions of $n_1$(heavy line), $n_2$(thin line),
$n_3$(dashed line) and $ \theta_1$(heavy line), $\theta_3$(dashed
line) in symmetric wells for strong interactions($c/v=5$),
 with initial values
(a)same to Fig.\ref{alla},
($n_1=0.4,n_3=0.05,\theta_1=0,\theta_3=0$); (b)
($n_1=0.49,n_3=0.012,\theta_1=0,\theta_3=2.82$);
(c)($n_1=0.15,n_3=0,\theta_1=0,\theta_3=0$);
(d)($n_1=0.076,n_3=0.066,\theta_1=2.54,\theta_3=2.79$);
(e)($n_1=0.96,n_3=0.02,\theta_1=1.36,\theta_3=1.34$);
(f)($n_1=0.348,n_3=0.532,\theta_1=0.817,\theta_3=3.843$).}
\label{all}
\end{figure*}

When the interaction is strong, the nonlinear effect is dominant,
accordingly the Poincar\'{e} section is complicated, as shown in
Fig.\ref{poinbi}, where we see many chaotic region except for some
stable islands. In the islands, the motions are periodic or
quasi-periodic whereas in the chaotic region the motions are
irregular. In order to grasp the dynamical property in this
situation, we have simulated the motions  for every regular islands
numerically, and find that except for the  oscillations like the
linear or weak interaction case, as shown in Fig.\ref{all}(a), there
are also four types of  new oscillations, as shown in
Fig.\ref{all}(b)--\ref{all}(e), respectively.

Fig.\ref{all}(a) shows that the oscillation in well one is almost
the same as  that of well three except a phase delay of
half-period. In order to compare with the linear or weak
interaction case, we take the same initial value as that of
Fig.\ref{alla}. We see that their oscillations behavior is
similar.

In addition to the  case shown in Fig.\ref{all}(a), the motions of
BECs in triple-well  can demonstrate  very different behavior.
Fig.\ref{all}(b) shows that almost all of BECs atoms oscillate
with small amplitude in two adjacent wells, i.e., well one and
well two. The phase $\theta_1$ oscillates around $0$ and the phase
$\theta_3$ oscillates around $\pi$. The energy of these
oscillations is closed to the eigen-energy of level labeled by
$E_{2a}$, and the center they surrounded is near the fixed point
corresponding to level $E_{2a}$. As mentioned before, this fixed
point are stable point.

Fig.\ref{all}(c) shows that almost all of BECs atoms oscillate
with large amplitude in well one and well two, and the relative
phase $\theta_1$ is always oscillating around zero.  These
oscillations are regarded as oscillations in a reduced two-well
trapped system. In fact, because $n_3$ is small and $c/v$ is very
large, we can regard the term
\begin{eqnarray}
{H_1}=v\sqrt{1-n_1-n_3}\sqrt{n_3}\cos\left(\theta_3\right)
\end{eqnarray}
as a perturbation. Using the generating function
\begin{eqnarray}
G=vg\left(J_1,J_2\right) \sin\left(\theta_3\right)+J_1 \theta _1+J_2\theta_3.
\end{eqnarray}
Where $$g\left(J_1,J_2\right)=\frac{v \sqrt{1-J_1-J_2}
   \sqrt{J_2}}{c \left(1-J_1-2
   J_2\right)}. $$
Then the Hamiltonian becomes
\begin{eqnarray}
H' &=& \frac{1}{2} c \left(2 J_1^2+2
   J_2 J_1-2 J_1+2 J_2^2-2
   J_2+1\right)                                     \nonumber \\
   && + v \sqrt{1-J_1-J_2}\sqrt{J_1}\cos
   \left(\Phi _1\right).
\end{eqnarray}
The new canonical variables have relations with the old canonical variables

\begin{subequations}
\begin{align}
 n_1 &= J_1                                           ,      \\
 n_3 &= J_2+v \cos \left(\theta _3\right)
   g\left(J_1,J_2\right)                                ,     \\
 \Phi _1 &= \theta _1+v \sin \left(\theta _3\right)
   g^{(1,0)}\left(J_1,J_2\right)                         ,    \\
 \Phi _2 &= \theta _2+v \sin \left(\theta _3\right)
   g^{(0,1)}\left(J_1,J_2\right)                            .
\end{align}
\label{newcano}
\end{subequations}

Since the action variable $J_2$ is constant,
the action-angle variables $J_1, \Phi_1$
can be solved first from the new canonical equations (\ref{newcano}).
Notice that $J_1=n_1$ is the population
in well one, $\Phi_1\approx \theta_1$ is the relative phase of quantum state in
well one and well two.
The oscillations shown in Fig.\ref{all}(c) just like the zero-phase
mode oscillations in a two-well trapped system.

Fig.\ref{all}(d) and \ref{all}(e) show self-trapping of BECs in
the middle well and self-trapping in one side well respectively.
These motions have a close relations to the property of fixed
points, and will be discussed in detail at the next section.

Fig.\ref{all}(f) shows the chaotic trajectory which corresponds to
the chaotic region in the Poincar\'{e} section. In this case, the
population in each well shows irregular oscillation with respect to
time.

\begin{figure}[t]
\centering
\includegraphics[width=0.5\textwidth]{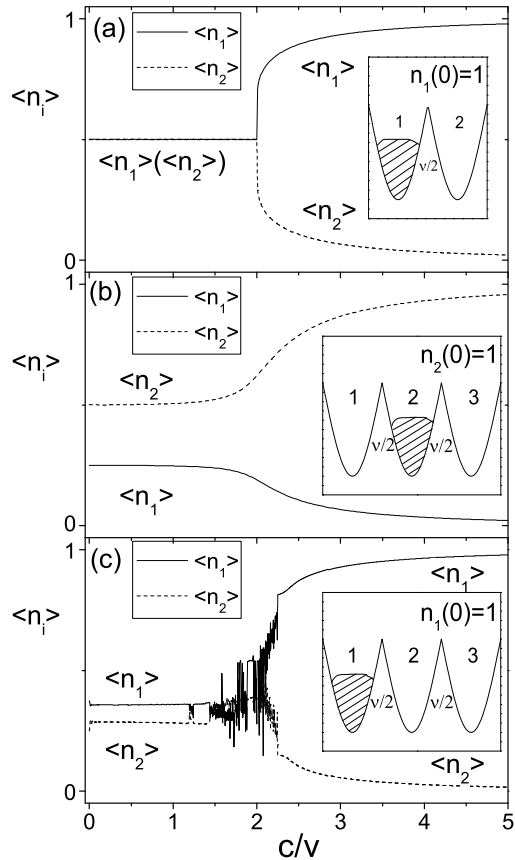}
\caption{The average of $n_1,n_2$ for different interactions $c/v$ in
two-well system(a) and triple-well system(b,c),
with initial value (a)$n_1(0)=1$, (b)$n_2(0)=1$, (c)$n_1(0)=1$. The schematic sketch
of the potential is shown in the figures.}
\label{selftrap}
\end{figure}

\section{Transition to Self-Trapping}

\subsection{Self-Trapping in One Well}

Self-trapping is caused by the nonlinear interactions. For symmetric
two-well system, as we have known, self-trapping happens only when
the interaction parameter excesses a critical value. By calculating
the averaged population for same initial value $a_1(0)=1,a_2(0)=0$
with different interaction strength $c/v$, we show this in
Fig.\ref{selftrap}(a). The Hamiltonian used for calculation is
$\hat{H}=\left(
\begin{array}{cc}
  c|a_1|^2 & -\frac{v}{2} \\
 -\frac{v}{2} & c|a_2|^2
 \end{array}
\right) $. It is clearly shown that, when $c/v >2$ the averaged
$n_1$ is no  longer zero indicating the beginning of the
self-trapping. Soon after that, with increasing the interaction
the BECs will be trapped in well one completely. Because the
transition corresponds to crossing over a separatrix from
oscillation to liberation, at the transition point $c/v=2$ the
scaling law follows a logarithm function\cite{wgf,logri}.

For our triple-well system, the high-dimensional phase space
permits the existence of chaos and the smooth movement of the
fixed points to the boundary ( the latter point will be clearly
shown in Fig.9 ). So we expect that the transition to self-trapping
in triple-well system will show distinguished property from that
of double-well system. To demonstrate it, we calculate the time
averaged population for different interactions with the initial
conditions $n_2(0)=1$ and $n_1(0)=1$, denoting initial BECs
uploaded in middle well and left-hand well, respectively. The
results are shown in Fig.\ref{selftrap}(b),\ref{selftrap}(c).

For the linear case, i.e., $c=0$, in Fig.\ref{selftrap}(b),
substituting $(a_1=0,a_2=1,a_3=0)$ to (\ref{initial}), we can get
one set of $C_i$
$$C_1=0, C_2=-\frac{i}{\sqrt{2}}, C_3=\frac{\pi}{2}, C_4=1, C_5=0.$$
Then from Eq.(\ref{solu}) we have the averaged populations
$$\langle n_1 \rangle=\langle n_3 \rangle=|C_2|^2/2=\frac{1}{4}, \langle n_2 \rangle=|C_4|^2/2=\frac{1}{2}.$$
Likewise, in Fig.\ref{selftrap}(c), substituting
$(a_1=1,a_2=0,a_3=0)$ to (\ref{initial}), we can get one set of
$C_i$
$$C_1=\frac{1}{2}, C_2=\frac{1}{2}, C_3=0, C_4=-\frac{i}{\sqrt{2}}, C_5=\frac{\pi}{2}.$$
From Eq.(\ref{solu}) we have the averaged populations
$$\langle n_2 \rangle=|C_4|^2/2=\frac{1}{4},\langle n_1 \rangle=\langle n_3 \rangle
=\frac{1-\langle n_2 \rangle}{2}=\frac{3}{8}.$$

For the weak interaction, i.e., $c<c_1$, the averaged populations
are still similar to the linear case. However, when the interaction
strength is close to the critical point $c_1$, the averaged
populations changed dramatically. For Fig.\ref{selftrap}(b),  the
averaged population $\langle n_2 \rangle$ increases monotonically
and smoothly to unit. No scaling law are observed.  This is due to
the fact that in the 4D phase space the fixed point can move
smoothly to the  boundary without through any bifurcation. For
Fig.\ref{selftrap}(c), the averaged populations become turbulence
near the critical point, this is a result of chaotic trajectory in
the phase space. Meanwhile, $\langle n_1 \rangle$ and  $\langle n_3
\rangle$ are no longer equal. When the interaction is larger than
$2.25$, the averaged population $\langle n_1 \rangle$ becomes smooth
and tends to unit rapidly.

\begin{figure}[t]
\centering
\includegraphics[width=0.5\textwidth]{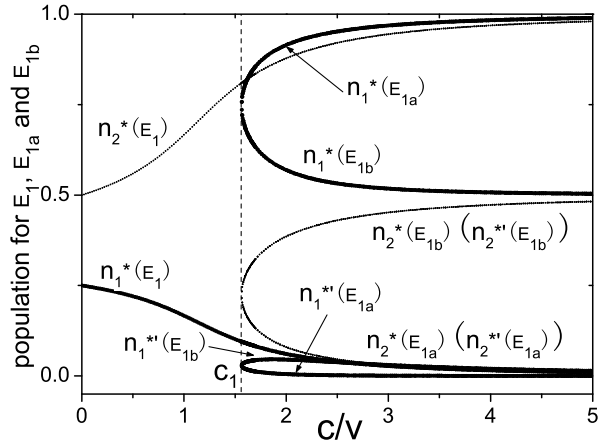}
\caption{The population $n_1^\ast$(heavy line) and $n_2^\ast$(thin line) for the levels labeled by
$E_1, E_{1a}$ and $E_{1b}$ in Fig.\ref{eigen}.}
\label{state}
\end{figure}

For triple-well system, when the interaction strength exceeds the
critical value $c_1$, as mentioned before, new levels $E_{1a},
E_{1b}$ will appear, accordingly new fixed points will emerge, and
the phase space will tend to be divided into several  subspace
around the stable fixed points. In Fig.\ref{state} we plot
populations $n_1^\ast$ and $n_2^\ast$  for the levels labeled by
$E_1,E_{1a}$ and $E_{1b}$ as a function of $c/v$. Corresponding to
eigenenergy  $E_{1a}$ there are two eigen-states, denoted by
$n_i^\ast$ and ${n_i^{\ast}} '$ respectively. The same thing
happens to level $E_{1b}$. Recall that $E_1,E_{1a}$ are stable
levels. When the interaction is very strong, the averaged
population $\langle n_2 \rangle$ of  the motions of BECs uploaded
initially in the middle well is
 close to $n_2^\ast(E_1)$, while
$\langle n_1 \rangle$ of  the motions of BECs uploaded initially
in well one is close to $n_1^\ast(E_{1a})$, as shown in
Fig.\ref{selftrap}(b),Fig.\ref{selftrap}(c).

\subsection{Self-Trapping in Two Wells}
In the above, we have investigated the self-trapping of BECs in
single  well. In this part, we will investigate  whether the BECs
atoms can  be self-trapped only in two wells.

\begin{figure}[t]
\centering
\includegraphics[width=0.48\textwidth]{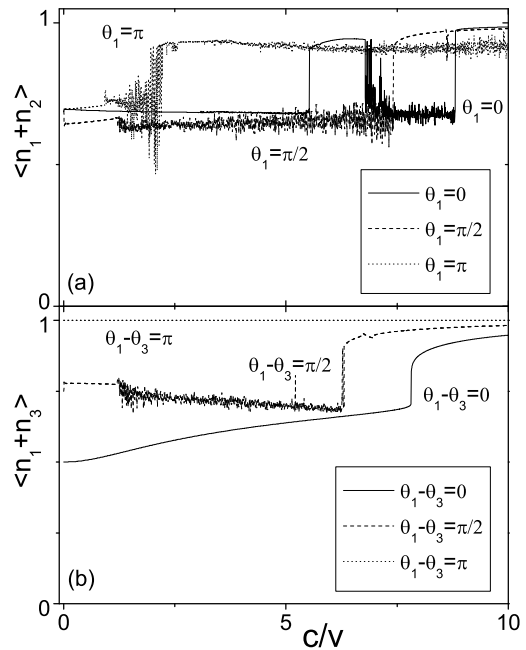}
\caption{The mean value of (a)$\langle n_1+n_2 \rangle$ and
(b)$\langle n_1+n_3 \rangle$ with initial value
(a)$(n_1=0.5,n_2=0.5,n_3=0)$, (b)$(n_1=0.5,n_2=0,n_3=0.5)$ for
different relative phase.} \label{twosel}
\end{figure}

Considering the interference, the initial relative phase should be
very important. So we calculate the mean value of $\langle n_1+n_2
\rangle$ as a function of $c/v$ for different phase $\theta_1$
with initial value $(n_1=0.5,n_2=0.5,n_3=0)$, and the mean value
of $ \langle n_1+n_3 \rangle$ as a function of $c/v$ for different
phase $\theta_1-\theta_3$ with initial value
$(n_1=0.5,n_2=0,n_3=0.5)$, respectively. The main results are
shown in Fig.\ref{twosel}.

It is shown that, with increasing the atomic interaction BECs will
be trapped in the two wells where it is initially uploaded and the
Josephson oscillation can be completely blocked. Both cases also
suggest that  the relative phase can dramatically influence the
transition to self-trapping for BECs. In Fig.\ref{twosel}(a), when
relative phase is zero, the Josephson oscillation and the
self-trapping can emerge alternately. Whereas, for case of
Fig.\ref{twosel}(b), $\pi$ value of the relative phase gives a
robust self-trapped BECs. In both cases, we also see the occurrence
of the chaos making the curves look irregular . Interestingly, the
onset of chaos can also be controlled by the relative phase, e.g.,
the vanishment of the relative phase will reduce the chaos and make
the BECs safely turn from oscillation state to self-trapping state
as shown in Fig.\ref{twosel}(b).

\section{Tilted Triple-Well, $\gamma\ne 0$}
In this section we extend the above discussions to the tilted
triple-well system. In this system, the diverse type of  Josephson
oscillations also emerge around the eigenstates. In light of the
energy spectra Fig.2, we can readily find out the parameter regime
for different kinds of oscillations. So, we will focus on the
transition to self-trapping that is of more interest. We want to see
how the transition is influenced by tilting the wells.
\subsection{Linear Oscillation Solutions}

For the linear case, i.e., $c=0$, the system is analytically
solvable. So the solutions of $(a_1,a_2,a_3)$ are

\begin{subequations}
\begin{align}
a_1 &= \frac{1}{2}C_4 \cos (\sqrt{\frac{v^2}{2}+\gamma^2}t+C_5)
 +\frac{1}{2(\frac{v^2}{2}+\gamma^2)}                          \nonumber          \\
& ( \gamma C_2 \cos (\sqrt{\frac{v^2}{2}+\gamma^2}t+C_3)
+v C_1 )                                                              ,           \\
a_2 &= \frac{1}{\frac{v^2}{2}+\gamma^2}(-\frac{v}{2}C_2
\cos (\sqrt{\frac{v^2}{2}+\gamma^2}t+C_3)                      \nonumber          \\
&+\gamma C_1)                                                         ,           \\
a_3 &=  \frac{1}{2} C_4 \cos (\sqrt{\frac{v^2}{2}+\gamma^2}t+C_5)
  -\frac{1}{2(\frac{v^2}{2}+\gamma^2)}                         \nonumber          \\
& (\gamma C_2 \cos (\sqrt{\frac{v^2}{2}+\gamma^2}t+C_3)+v C_1)  .
\end{align}
\label{asolu}
\end{subequations}

Where $C_i$ are complex integral constant and like the symmetric case, determined by
initial conditions.

It is clear that $a_1,a_2$ and $a_3$ vary with respect to time
periodically. The period $T=2\pi/\sqrt{\frac{v^2}{2}+\gamma^2}$.

\subsection{Transition to Self-Trapping}

\begin{figure}[t]
\centering
\includegraphics[width=0.5\textwidth]{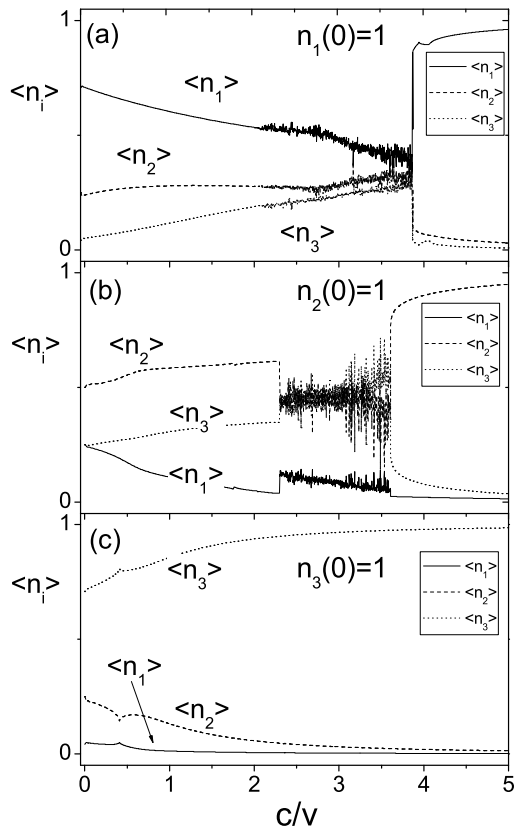}
\caption{The average of $n_i$ for different interactions $c/v$ in asymmetric wells $\gamma=-1$
with initial value (a)$n_1(0)=1$, (b)$n_2(0)=1$, (c)$n_3(0)=1$.}
\label{selftrapt}
\end{figure}

 When the interaction is  strong we still observe the self-trapping of BECs in the tilted system.
 We plot  the averaged
populations in their dependence of the interaction strength with
choosing parameter $\gamma=-1$ and  initial conditions $n_1=1$,
$n_2=1$ and $n_3=1$ in Fig.\ref{selftrapt}(a),\ref{selftrapt}(b) and
 \ref{selftrapt}(c)  respectively.
The schematic sketch of the wells is shown in Fig.\ref{potential}(b).

For Fig.\ref{selftrapt}(a), when $c/v=0$,
with the initial conditions $(a_1=1,a_2=0,a_3=0)$,
we derive one set of $C_i$

\begin{eqnarray}
&& C_1=v/2, C_2=i v/\sqrt{2}, C_3=\arccos(\frac{i\sqrt{2}\gamma}{v}) ,\nonumber \\
&&  C_4=\frac{v}{\sqrt{v^2+2\gamma^2}},
 C_5=-\arccos(\frac{\sqrt{v^2+2\gamma^2}}{v}).\nonumber
\end{eqnarray}

By integrating $|a_i(t)|^2$ with respect to time and making average
over time we obtain the averaged population analytically from (\ref{asolu}),
$$\langle n_1 \rangle=17/24, \langle n_2 \rangle=1/4, \langle n_3 \rangle=1/24 .$$

With increasing the nonlinearity, the averaged population in well
one $\langle n_1 \rangle$ decrease at first, and then become
turbulence, indicating   chaotic motions. When the interaction
parameter is larger than $3.9$, the averaged population jumps up and
tends to unit soon after.

In Fig.\ref{selftrapt}(b),  initial condition is
$(a_1=0,a_2=1,a_3=0)$. Similarly the  averaged populations are
readily obtained for the linear case, $\langle n_1 \rangle=1/4,
\langle n_2 \rangle=1/2, \langle n_3 \rangle=1/4 .$ With increasing
the nonlinearity, the averaged population $\langle n_2 \rangle$
increases smoothly at beginning, passes a turbulence interval
$[2.3,3.6]$, and then  jumps up  and tends to unit. Another
interesting phenomenon in the process is that the averaged
population in well three is always larger than that of well one in
the presence of the nonlinearity, even though in this case the
zero-point energy of well three is obviously  greater than that of
well one. This somehow counterintuitive phenomenon is clearly the
consequence of the nonlinearity.

In Fig.\ref{selftrapt}(c), when $c/v=0$, with the initial conditions
$(a_1=0,a_2=0,a_3=1)$, the averaged populations have some
correspondence to  that of  Fig.\ref{selftrapt}(a) due to behind
symmetry, i.e., $\langle n_1 \rangle=1/24, \langle n_2 \rangle=1/4,
\langle n_3 \rangle=17/24 .$ With increasing the nonlinearity,  the
averaged population in well three $\langle n_3 \rangle$ increases
monotonically and smoothly to unit.

With comparing Fig.\ref{selftrapt} to Fig.\ref{selftrap}, we find that, the smooth transition
of the BECs in middle well to self-trapping in the symmetric
triple-well is broken by tilting the wells, and lifting the well
three makes BECs smoothly transit to self-trapping states without
losing their stability. Fig.11 also shows that, BECs in higher wells
are easily self-trapped. From the above discussions, we conclude
that the transition to self-trapping of BECs in triple-well systems
can be effectively controlled by tilting the wells.

\section{Conclusions}
We have presented a  comprehensive analysis of the tunnelling
dynamics for BECs in a triple-well trap both numerically and
analytically. Diverse energy levels are demonstrated. Behind these
unusual level structures, we reveal many new types of nonlinear
Josephson oscillation. We also study the self-trapping of BECs in
one-well as well as in the two-well and  investigate the
transition from nonlinear Josephson oscillation  to
self-trappings. Distinguished from the double-well case, no
scaling law is observed at the transition and the transition may
be accompanied by a irregular regime where the motions are
dominated by chaos. We also find that the transition can be
effectively controlled by the relative phase between wells and
tilting the wells.
 In the present experiments, the
double-well is realized in the optical traps with using a
blue-detuned light to form a barrier.  With the same technique, the
triple-well is also possibly realized in the optical traps. We hope
our theoretical discussion will stimulate the experiments in the
direction.

\section*{Acknowledgement}

This work was supported by National Natural Science Foundation of
China (No.10474008,10604009), Science and Technology fund of CAEP,
the National Fundamental Research Programme of China under Grant No.
2005CB3724503, the National High Technology Research and Development
Program of China (863 Program) international cooperation program
under Grant No.2004AA1Z1220, and Hebei Natural Science Foundation
Project under Grant No.: A2006000128.


\begin{thebibliography}{99}

\bibitem [*] {} Liu$\_$Jie@iapcm.ac.cn

\bibitem{first}
        M. H. Anderson, M. R. Matthews, C. E. Wieman, and
    E. A. Cornell, Science \textbf{269}, 198 (1995);
        K. B. Davis, M. -O. Mewes, M. R. Andrews, N. J. van Druten,
    D. S. Durfee, D. M. Kurn, and W. Ketterle, Phys. Rev. Lett. \textbf{75}, 3969 (1995);
        C. C. Bradley, C. A. Sackett, J. J. Tollett and R. G.
    Hulet, \textit{ibid}. \textbf{75}, 1687 (1995).

\bibitem{gpe}
        L. Pitaevskii and S. Stringari, Bose-Einstein Condensation (Oxford
    University Press, Oxford, 2003).

\bibitem{twomode}
        A.Smerzi, S. Fantoni, S. Giovanazzi, and S. R. Shenoy, Phys. Rev.
    Lett. \textbf{79}, 4950 (1997);
        G. J. Milburn, J. Cnrney, E. M. Wright, and D. F. Walls, Phys. Rev. A \textbf{55},
    4318 (1997).

\bibitem{niu}
        Biao Wu and Qian Niu, Phys. Rev. A \textbf{61}, 023402 (2000);
        O. Zobay and B. M. Garraway, Phys. Rev. A \textbf{61}, 033603 (2000);
        R. D' Agosta and C. Presilla, Phys. Rev. A \textbf{65}, 043609 (2002).

\bibitem{stat}
        M. Holthaus, Phys. Rev. A \textbf{64}, 011601(R) (2001).

\bibitem{jose}
        S. Raghavan, A. Smerzi, S. Fantoni, and S. R. Shenoy, Phys.
    Rev. A \textbf{59}, 620 (1999).

\bibitem{lzt}
        Jie Liu, Libin Fu, Bi-Yiao Ou, Shi-Gang Chen, Dae-II Choi, Biao Wu, and Qian Niu,
    Phys. Rev. A \textbf{66}, 023404 (2002).

\bibitem{leggett}
        Anthony J. Leggett, Rev. Mod. Phys. \textbf{73}, 307 (2001),
    and references therein.

\bibitem{add}
        D. Witthaut, E. M. Graefe, and H. J. Korsch,,Phys. Rev. A \textbf{73}, 063609
        (2006);
        Biao Wu and Jie Liu, Phys. Rev. Lett. \textbf{96}, 020405
        (2006).
\bibitem{qm}
        L. D. Landau and E. M. Lifshitz, Quantum Mechanics (Pergamon Press, New York, 1997).

\bibitem{exper}
        Michael Albiez, R. Gati, Jonas F\"{o}lling, S. Hunsmann, M.
    Cristiani, and M. K. Oberthaler, Phys. Rev. Lett. \textbf{95}, 010402 (2005).

\bibitem{tri}
        E. M. Graefe, H. J. Korsh, and D. Witthaut, Phys. Rev. A \textbf{73}, 013617 (2006).

\bibitem{ple}
        S. Mossmann and C. Jung, Phys. Rev. A \textbf{74}, 033601 (2006).

\bibitem{bbw}
        The "self-trapping" of BECs in optical lattice has been observed
    experimentally, however the behind physics is not fully understood,
    refer to,
        T. Anker, M. Albiez, R. Gati, S. Hunsmann, B.
    Eiermann, A. Trombettoni, and M. K. Oberthaler, Phys. Rev. Lett. 94,
    020403 (2005);
        Tristram J. Alexander, Elena A. Ostrovskaya, and Yuri
    S. Kivshar,Phys. Rev. Lett. \textbf{96}, 040401 (2006);
        Roberto Livi, Roberto Franzosi, and Gian-Luca Oppo,Phys. Rev.
    Lett. \textbf{97}, 060401 (2006);
        Bingbing Wang, Panming Fu, Jie Liu, and
    Biao Wu, e-print cond-mat/0601249 (2006).

\bibitem{ps}
        A. J. Lichtenberg, and M. A. Lieberman, Regular and Stochastic
    Motion (Springer-Verlag, New York, 1983);
        L. E. Reichl, The Transition to Chaos (Springer-Verlag, New York, 1992).


\bibitem{clas}
        Jie Liu, Biao Wu, and Qian Niu, Phys. Rev. Lett. \textbf{90}, 170404 (2003).

\bibitem{wolf}
        An introduction of Mathematica could be found from
    \href{http}{http://www.wolfram.com/}.

\bibitem{adda}
        Jie Liu, Chuanwei Zhang, Mark G. Raizen, and Qian
    Niu, Phys. Rev. A \textbf{73}, 013601 (2006);
        Chuanwei Zhang, Jie Liu, Mark G. Raizen, and Qian Niu,
    Phys. Rev. Lett. \textbf{92}, 054101 (2004).


\bibitem{wgf}
        Guan-Fang Wang, Li-Bin Fu, and Jie Liu, Phys. Rev. A \textbf{73}, 013619 (2006).

\bibitem{logri}
        Li-Bin Fu and Jie Liu, eprint cond-mat/0609337 (2006).

\end{thebibliography}
\end{document}